\documentclass{article}
\usepackage[utf8]{inputenc}
\usepackage{amsmath}
\usepackage{hyperref}
\usepackage{graphicx}
\usepackage{tikz}
\usepackage{pgfplots}
\usepackage{booktabs}
\usepackage{abstract}

\title{\textbf{Encoding sinusoidal functions in hybrid automata formalism}}
\author{\textbf{Nikolaos Kekatos}\\ \emph{Verimag Laboratory, University of Grenoble Alpes}\\
 \href{mailto:nikolaos.kekatos@univ-grenoble-alpes.fr}{nikolaos.kekatos@univ-grenoble-alpes.fr}
 }
\date{}

\begin{document}

\maketitle
\begin{abstract}
    Hybrid systems can express a plethora of physical phenomena and systems as they can combine continuous and discrete dynamics. There exist several tools which enable the reachability analysis of hybrid systems modeled as hybrid automata. However, these tools exhibit certain limitations in the type of mathematical operations that they natively support. For example, SpaceEx, a well established tool in the hybrid verification community, supports the use of linear ODEs in the flow of each discrete location. Mathematical functions like alegbraic equations or trigonometric functions have to be encoded as the solutions of a set of ODEs. In this article, we provide a mechanism to define sinusoidal functions that is supported by SpaceEx. We also note how certain Simulink blocks can be translated in hybrid automata.
\end{abstract}
\section{Introduction}

Herein, we illustrate how we can use linear ODEs to express sinusoids and how to describe them as hybrid automata. Note that our proposed scheme is suitable for sine waves (functions of time) and applies to formal verification tools, such as SpaceEx~\cite{Frehse2011}.

\section{Sinusoids via linear ODEs}\label{sec2}

The differential equation
\begin{align}
\begin{aligned}
\dot{x} &= y \\
\dot{y} &= -\omega^2 (x-\mu t)
\end{aligned}
\label{eq1}
\end{align}
where $x$, $y$ are the state variables, $\omega$ and $\mu$ are parameters, and $t$ describes the time evolution, has the following analytical solution
\begin{subequations}
\begin{align}
y(t) &= A \sin(\omega t +\varphi) + \mu \\ 
x(t) &= -\frac{A}{\omega}\cos(\omega t + \varphi) + \mu t.
\end{align}
\end{subequations}
The initial conditions are 
$y(0)=A\sin(\varphi) + \mu$, and $x(0) =  -\frac{A}{\omega}\cos(\varphi)$. If we ignore the initial phase, i.e. $\varphi=0$, the initial conditions become $y(0) = \mu$, $x(0) = -\frac{A}{\omega}$.

\section{Sinusoids as hybrid automata}

\begin{figure}[ht!]
\centering
\includegraphics[scale=0.5]{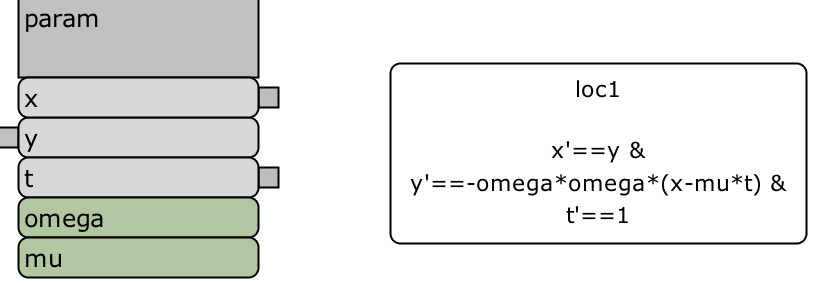}
\caption{Sinusoid encoded as a base component of a hybrid automaton (depicted in SpaceEx MO.E.)}
\label{sin_sx}
\end{figure}

Hybrid automata are a common formalism for modeling hybrid systems and their semantics can be found at~\cite{frehse2015introduction}. In this part, we show how to construct hybrid automata that respect the SX format~\cite{cotton2010spaceex}; a formalism that supports network of hybrid automata and is used by several reachability tools, such as SpaceEx. ~\\
\noindent \textbf{Base Component - Sine.} For the sinusoid, we only need one \emph{base component} that we call {\sf{sin}}.
We have one \emph{location}, named {\sf{loc1}}, with three variables $x$, $y$, and $t$. Using the equation~\ref{eq1}, the {\emph{flow}} is given by\footnote{Note that we opt for $omega*omega$ instead of $\omega^2$  to avoid any unexpected error caused by the SpacEx parser.}
\begin{align*}
x' &== y \\
y' &== -\omega*\omega *(x-\mu * t)\\
t'&==1
\end{align*}
The three variables $x$, $y$, $t$ are declared as controlled, while the parameters~/ constants  $\omega$ and $\mu$ are uncontrolled. The variable $y$ is a global variable, i.e. it can be communicated/shared to other components, whereas $y$ and $t$ are local variables. The hybrid automaton requires no invariant, reset, guard or synchronization label. Detailed information about modeling hybrid automata in SpaceEx can be found at~\cite{SpaceEX}. The complete base component is shown in Figure~\ref{sin_sx}.

The initial conditions are selected according to Section~\ref{sec2} to match the desired behavior of the sinusoidal signal.\\

\begin{figure}[ht!]
\centering
\includegraphics[scale=0.55]{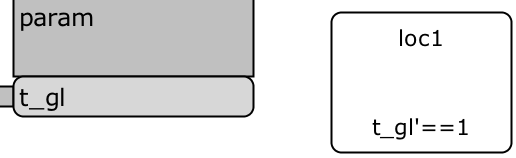}
\caption{Clock encoded as a base component of a hybrid automaton (depicted in SpaceEx MO.E.)}
\label{clock}
\end{figure}

\noindent \textbf{Base Component - Clock.}
We also need a base component that measures the global time. The name of the base component is {\sf clock} and it has one variable $t_{gl}$. The variable is controlled and non-local. The base component has only one location with flow $\dot{t}_{gl}=1$. The base component is portrayed in Figure~\ref{clock}.%

\begin{figure}[ht!]
\centering
\includegraphics[scale=0.55]{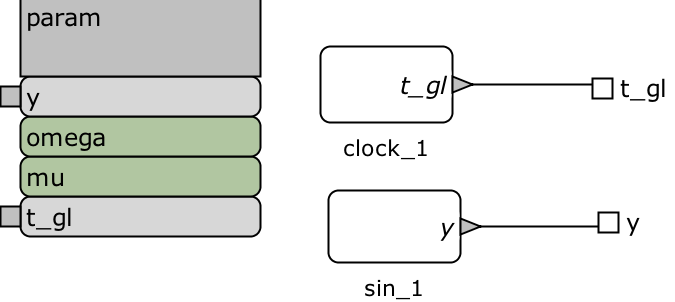}
\caption{Network component containing a sinusoid and a clock base component  (depicted in SpaceEx MO.E.)}
\label{network}
\end{figure}

%
\noindent \textbf{Network Component.} For testing purposes, we add a network component that is a network of two hybrid automata. To do so, we add two binds in the Model Editor. The user can select the values of the parameters $\omega$ and $\mu$ here. The network component is portrayed in Figure~\ref{network}.%

\section{Application}

In this section, we show the applicability of this transformation via a numerical example as well as its usability within Simulink.
\subsection{Numerical Example}

\begin{figure}[ht!]
\centering
\includegraphics[scale=0.55]{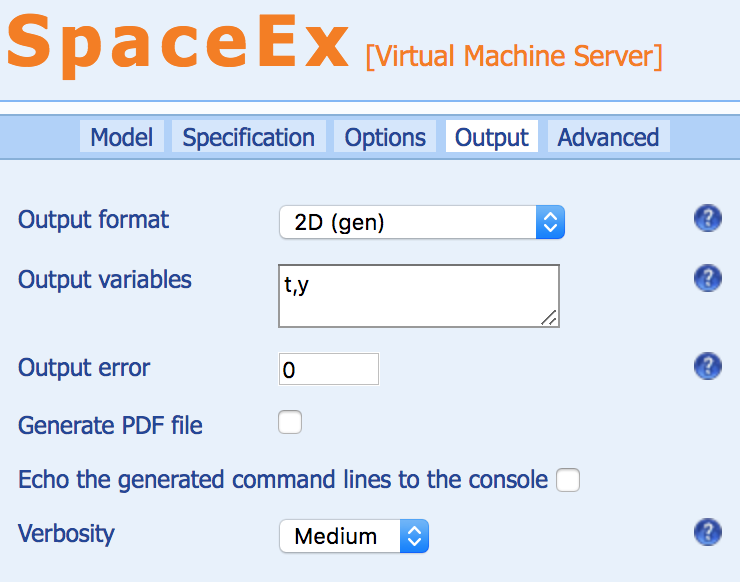}
\caption{Selecting the output states in SpaceEx web interface}
\label{out}
\end{figure}

\begin{figure}[ht!]
\centering
\includegraphics[scale=0.55]{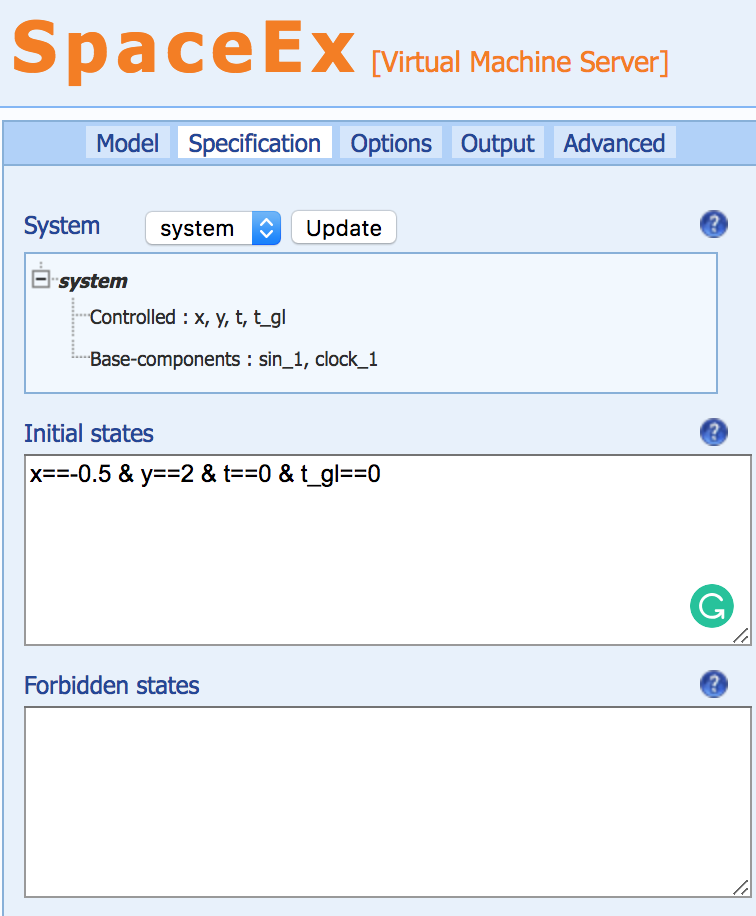}
\caption{Selecting the initial conditions in SpaceEx web interface}
\label{spec}
\end{figure}

Consider that we opt for $\omega=1$, $\mu=2$, $A=0.5$, and $\phi=0$. After loading the model in the web interface, we need to select the initial states and the output variables. Note that $x(0)=-\frac{A}{\omega}\cos(\phi)=-0.5$, $y(0)=A\sin(\varphi) + \mu=2$, and $t(0)=0$. For SpaceEx use, the output variables are defined in the {\sf{Output}} tab, see Figure~\ref{out} and the initial states in the {\sf Specification} tab, see Figure~\ref{spec}.

\begin{figure}[ht!]
\centering
\scalebox{0.95}{
\begin{tikzpicture}
\begin{axis}
\addplot [color=red,fill=red!90!white,opacity=0.5] table {reach.gen};
\end{axis}
\end{tikzpicture}}
\caption{Reachability analysis with SpaceEx web interface (plotted with TikZ).}
\label{reach}
\end{figure}
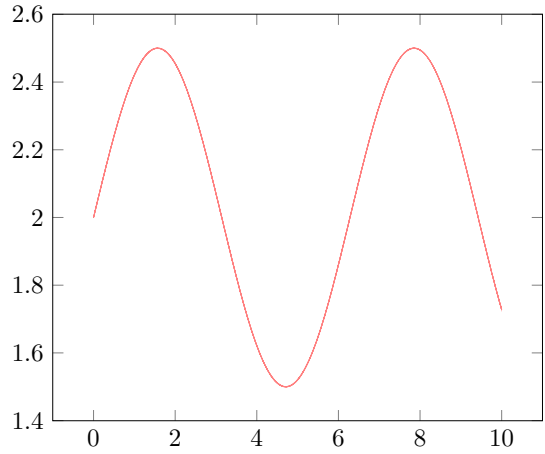

Then, we finalize the reachability analysis features in the {\sf Options} tab by selecting the STC scenario, flowpipe tolerance 0.01, local time horizon 10s and max. iterations -1 (fixed point). The reachable set is shown in Figure~\ref{reach}.  

\subsection{Simulink sine wave block}

\begin{figure}[ht!]
\centering
\includegraphics[scale=0.55]{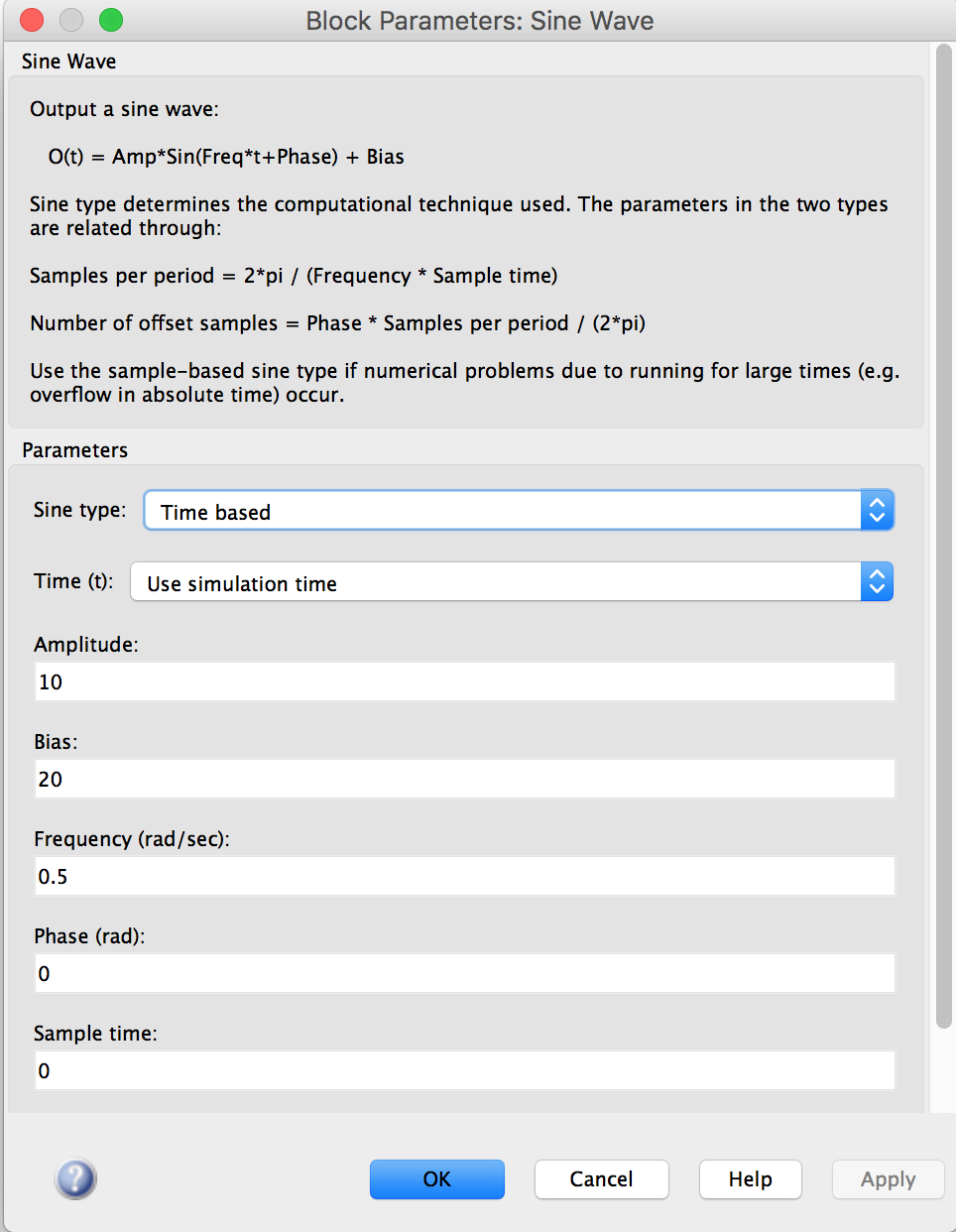}
\caption{Screenshot of a sine wave block in Simulink}
\label{simulink}
\end{figure}

The proposed translation scheme is directly applicable to the Simulink sine wave block\footnote{\url{https://www.mathworks.com/help/simulink/slref/sinewave.html}}. Figure~\ref{simulink} shows the options that exist in such a Simulink block. Table~\ref{table} presents the relation between the options in Simulink and the corresponding variables in SpaceEx. The translation of some Simulink blocks into hybrid automata supported by SpaceEx formalism is shown in~\cite{kekatos2017constructing,kekatos2018formal}.

\begin{table}[ht!]
\centering
\caption{Translating Simulink parameters into SpaceEx parameters\label{table}}
\begin{tabular}{lc}
 \toprule
 Simulink & SpaceEx \\
 \midrule
 Amplitude & $A$\\
 Bias & $\mu$\\
 Frequency & $\omega$\\
 Phase & $\phi$ \\
 Sample time & --\\
\bottomrule
\end{tabular}
\end{table}
As it can be observed, the user can directly input the values of the Simulink block into the corresponding SpaceEx parameters. The initial conditions should be selected accordingly. Note that we shall opt the sampling time to be 0 as handle continuous-time systems. 

\begin{figure}[ht!]
\centering
\includegraphics[scale=0.5]{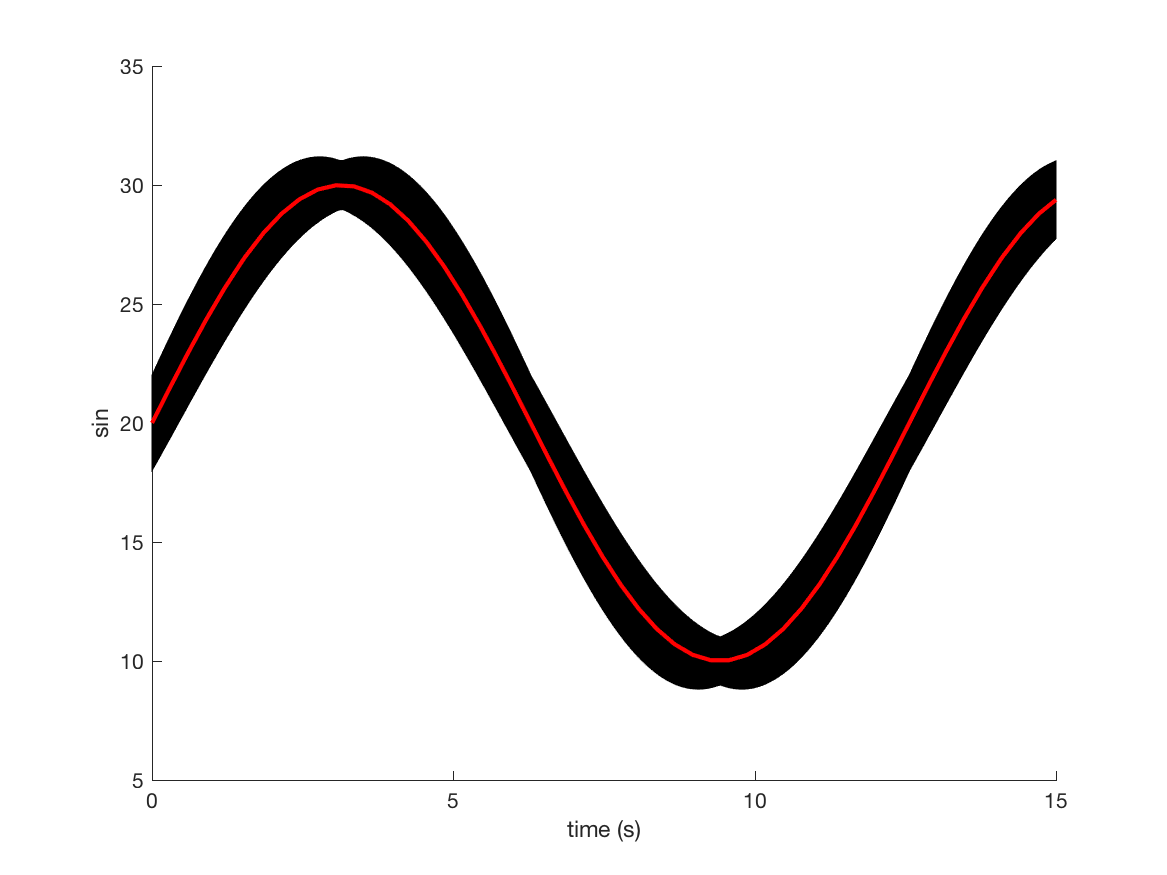}
\caption{Simulation results (Simulink) in red vs Reachable sets (SpaceEx) in black. Plot done in MATLAB.}
\label{comparison_sin}
\end{figure}

\newpage
For the aforementioned example, we have $A=10$, $\mu=20$, $\omega=0.5$, and $\phi=0$. Then, $x(0)=-20$ and $y(0)=20$. Figure~\ref{comparison_sin} shows the reachable set computed with SpaceEx against the Simulink simulation. The initial conditions are enlarged by 20\% for the reachability computations. It can be observed that the simulation is included in the flowpipe.

\newpage
\bibliographystyle{plain}
\bibliography{references.bib}

\begin{thebibliography}{1}

\bibitem{cotton2010spaceex}
Scott Cotton, Goran Frehse, and Olivier Lebeltel.
\newblock The {SpaceEx} modeling language.
\newblock
  \url{http://spaceex.imag.fr/sites/default/files/spaceex_modeling_language_0.pdf},
  2010.

\bibitem{SpaceEX}
Goran Frehse.
\newblock {An Introduction to {SpaceEx} v0.8}.
\newblock
  \url{http://spaceex.imag.fr/documentation/user-documentation/introduction-spaceex-27},
  2010.

\bibitem{frehse2015introduction}
Goran Frehse.
\newblock {An Introduction to Hybrid Automata, Numerical Simulation and
  Reachability Analysis}.
\newblock In {\em Formal Modeling and Verification of Cyber-Physical Systems},
  pages 50--81. Springer, 2015.

\bibitem{Frehse2011}
Goran Frehse, Colas Le~Guernic, Alexandre Donz{\'e}, Scott Cotton, Rajarshi
  Ray, Olivier Lebeltel, Rodolfo Ripado, Antoine Girard, Thao Dang, and Oded
  Maler.
\newblock {SpaceEx}: Scalable verification of hybrid systems.
\newblock In {\em {CAV Conference}}, 2011.

\bibitem{kekatos2018formal}
Nikolaos Kekatos.
\newblock {\em Formal Verification of Cyber-Physical Systems in the Industrial
  Model-Based Design Process}.
\newblock PhD thesis, 2018.

\bibitem{kekatos2017constructing}
Nikolaos Kekatos, Marcelo Forets, and Goran Frehse.
\newblock Constructing verification models of nonlinear simulink systems via
  syntactic hybridization.
\newblock In {\em 2017 IEEE 56th Annual Conference on Decision and Control
  (CDC)}, pages 1788--1795. IEEE, 2017.

\end{thebibliography}
\end{document}